\documentclass{PoS}
\usepackage{epsfig}
\usepackage{mcite}
\PoS{PoS(LAT2005)030}
\newcommand{\be}{\begin{equation}}
\newcommand{\ee}{\end{equation}}

\title{Density-density correlators using all-to-all propagators }

\ShortTitle{Density-density correlators using all-to-all propagators }

\author{Constantia Alexandrou\\
        Department of Physics, University of Cyprus, CY-1678 Nicosia, Cyprus\\
        E-mail: \email{alexand@ucy.ac.cy}}
\author{Petros Dimopoulos\\
        Department of Physics, University of Cyprus, CY-1678 Nicosia, Cyprus\\
        E-mail: \email{dimop@ucy.ac.cy}}
\author{\speaker{Giannis Koutsou}\\
        Department of Physics, University of Cyprus, CY-1678 Nicosia, Cyprus\\
        E-mail: \email{koutsou@ucy.ac.cy}}
\author{Hartmut Neff\\
        University College London, Center for Computational Science, 20 Gordon Street, London WC1, UK\\
        E-mail: \email{uccahne@ucl.ac.uk}}

\abstract{We present a study of gauge invariant density-density
correlators.  Density-density correlators probe hadron wave functions and
thus can be used to study hadron deformation. Their zero momentum projection requires the computation
of all-to-all propagators, which are evaluated with the standard
stochastic technique, the dilution method and the stochastic sequential technique. 
We compare the results to a previous analysis that did not employ the zero momentum projection.}

\FullConference{XXIIIrd International Symposium on Lattice Field Theory\\
		 25-30 July 2005\\
		 Trinity College, Dublin, Ireland}

\begin{document}
\section{Introduction}
The shape of hadrons is a fundamental property that depends on QCD dynamics. In
particular, determining the shape of the nucleon has been an open issue for decades and recently a number of experiments have been performed in order to look for deformation
in the nucleon system \cite{PRL:86-2963,*PRL:88-122001}. 
Lattice QCD provides a model independent method to study  hadron
deformation through hadron wave functions and quark distributions. 
In this work we study hadron deformation by evaluating density-density correlation
functions \cite{PRD:34-3882,*NPB:307-669,*AP:238-441,*NPPS:30-419}. In the non-relativistic limit they reduce to the wave function squared and
being gauge invariant quantities are preferable to Bethe - Salpeter amplitudes. 
\vspace*{-0.25cm}
\section{Definitions and setup}
The density-density correlation function for a meson is given by
\be
C\left(\vec{y},t_1,t_2\right)=\int d^3x \left<h\right|\rho^{f_1}\left(\vec{x}+\vec{y},t_1\right)\rho^{f_2}\left(\vec{x},t_2\right)\left|h\right>,
\label{Eq:DDcorr}
\ee
where $h$ represents a meson state and $f_1$,$f_2$ are flavor indices. It is shown schematically in Fig.~\ref{Fig:Meson}. The charge density
operator in normal ordering is given by
\be
\rho^f\left(\vec{x},t\right)=:\bar{q}_f\left(\vec{x},t\right)\gamma_0q_f\left(\vec{x},t\right):.
\label{Eq:Density}
\ee
In general, the corresponding baryon correlator has three density insertions on quark lines as shown in Fig.~\ref{Fig:Baryon}, 
two of which can be on the same quark line as shown in Fig.~\ref{Fig:Baryon}b. Integrating the correlator shown in Fig.~\ref{Fig:Baryon}a,
over one relative coordinate reproduces, to a good approximation, the density-density baryon correlator. The latter can be thought of as 
the square of the one-particle wave function and depends only on one relative distance \cite{PRD:66-094503,*PRD:68-074504,*NPPS:128-1}. 
Therefore information about baryon deformation is encoded in the one-particle
wave function and can be extracted by studying the density-density correlators described 
in Eq.~(\ref{Eq:DDcorr}) with $h$ being the appropriate baryonic
state. We consider a maximum time separation between the source and the sink of $T/2$
where $T$ denotes the time extent of the lattice with anti-periodic boundary conditions in the temporal direction. 
An optimal suppression of excited states can be achieved by
choosing the density insertions to be at a time separation of $T/4$ from the source and the sink
\cite{PRD:66-094503,*PRD:68-074504,*NPPS:128-1}. Summation over the sink coordinates projects to the
zero momentum state. This requires knowledge of the all-to-all propagator, making the computation a difficult numerical task.
\begin{figure}[h]
  \begin{minipage}[h]{0.325\linewidth}
    \centering
    \scalebox{0.2}{
      \begin{picture}(0,0)%
	\includegraphics{Meson.pstex}%
      \end{picture}%
      \setlength{\unitlength}{3947sp}%
      \begingroup\makeatletter\ifx\SetFigFont\undefined%
      \gdef\SetFigFont#1#2#3#4#5{%
	\fontsize{#1}{#2pt}%
	\fontfamily{#3}\fontseries{#4}\fontshape{#5}%
	\selectfont}%
      \fi\endgroup%
      \begin{picture}(10420,3944)(871,-5933)
	\put(6526,-2536){\makebox(0,0)[lb]{\smash{{\SetFigFont{20}{24.0}{\familydefault}{\mddefault}{\updefault}{$\rho^u\left(\vec{x},T/4\right)$}%
	}}}}
	\put(6526,-5686){\makebox(0,0)[lb]{\smash{{\SetFigFont{20}{24.0}{\familydefault}{\mddefault}{\updefault}{$\rho^d\left(\vec{x}+\vec{y},T/4\right)$}%
	}}}}
	\put(10876,-3736){\makebox(0,0)[lb]{\smash{{\SetFigFont{20}{24.0}{\familydefault}{\mddefault}{\updefault}{$0$}%
	}}}}
	\put(901,-3661){\makebox(0,0)[lb]{\smash{{\SetFigFont{20}{24.0}{\familydefault}{\mddefault}{\updefault}{$T/2$}%
	}}}}
      \end{picture}%
    }
    \caption{The density-density correlator for the meson. $T$ is the temporal length of the lattice used.}
    \label{Fig:Meson}
  \end{minipage}
  \hspace*{0.01\linewidth}
  \begin{minipage}[h]{0.6\linewidth}
    \begin{center}
      \parbox[t]{0.3\linewidth}{   
	\centering
	\scalebox{0.2}{
	  \begin{picture}(0,0)%
	    \includegraphics{BaryonA.pstex}%
	  \end{picture}%
	  \setlength{\unitlength}{3947sp}%
	  \begingroup\makeatletter\ifx\SetFigFont\undefined%
	  \gdef\SetFigFont#1#2#3#4#5{%
	    \reset@font\fontsize{#1}{#2pt}%
	    \fontfamily{#3}\fontseries{#4}\fontshape{#5}%
	    \selectfont}%
	  \fi\endgroup%
	  \begin{picture}(10465,4755)(826,-7312)
	    \put(826,-3661){\makebox(0,0)[lb]{\smash{{\SetFigFont{20}{24.0}{\familydefault}{\mddefault}{\updefault}{$T/2$}%
	    }}}}
	    \put(10876,-3736){\makebox(0,0)[lb]{\smash{{\SetFigFont{20}{24.0}{\familydefault}{\mddefault}{\updefault}{$0$}%
	    }}}}
	  \end{picture}%
	}
      }
      \hspace*{0.2\linewidth}
      \parbox[t]{0.3\linewidth}{
	\centering
	\scalebox{0.2}{
	  \begin{picture}(0,0)%
	    \includegraphics{BaryonB.pstex}%
	  \end{picture}%
	  \setlength{\unitlength}{3947sp}%
	  \begingroup\makeatletter\ifx\SetFigFont\undefined%
	  \gdef\SetFigFont#1#2#3#4#5{%
	    \reset@font\fontsize{#1}{#2pt}%
	    \fontfamily{#3}\fontseries{#4}\fontshape{#5}%
	    \selectfont}%
	  \fi\endgroup%
	  \begin{picture}(10465,4755)(826,-7312)
	    \put(826,-3661){\makebox(0,0)[lb]{\smash{{\SetFigFont{20}{24.0}{\familydefault}{\mddefault}{\updefault}{$T/2$}%
	    }}}}
	    \put(10876,-3736){\makebox(0,0)[lb]{\smash{{\SetFigFont{20}{24.0}{\familydefault}{\mddefault}{\updefault}{$0$}%
	    }}}}
	  \end{picture}%
	}
      }
    \end{center}
    \caption{The density-density correlator for the baryon. The symbol $(\times)$ denotes the density insertion.}
    \label{Fig:Baryon}
  \end{minipage}
\end{figure}

There are two ways to address this problem:
\begin{itemize}
\item[a)] As a first approach to the problem one could neglect the summation over the sink and ensure that the  
suppression of the non-zero momenta is obtained by the time separation $T/4$ \cite{PRD:66-094503,*PRD:68-074504,*NPPS:128-1}.
In the non-relativistic limit where the center mass wave function factorizes as $e^{i\vec{P}\cdot\vec{r}}$ the correlators are independent of the total momentum, $\vec{P}$.

\item[b)] One develops an efficient method to obtain the all-to-all propagator using stochastic noise techniques.
This enforces hadronic states with zero momentum. We will be referring to this method as the stochastic method.
\end{itemize}
\vspace*{-0.5cm}
\section{Density-density correlators without explicit zero momentum projection}
\begin{figure}[h]
\begin{minipage}[t]{0.45\linewidth}
\begin{center}
\epsfig{file=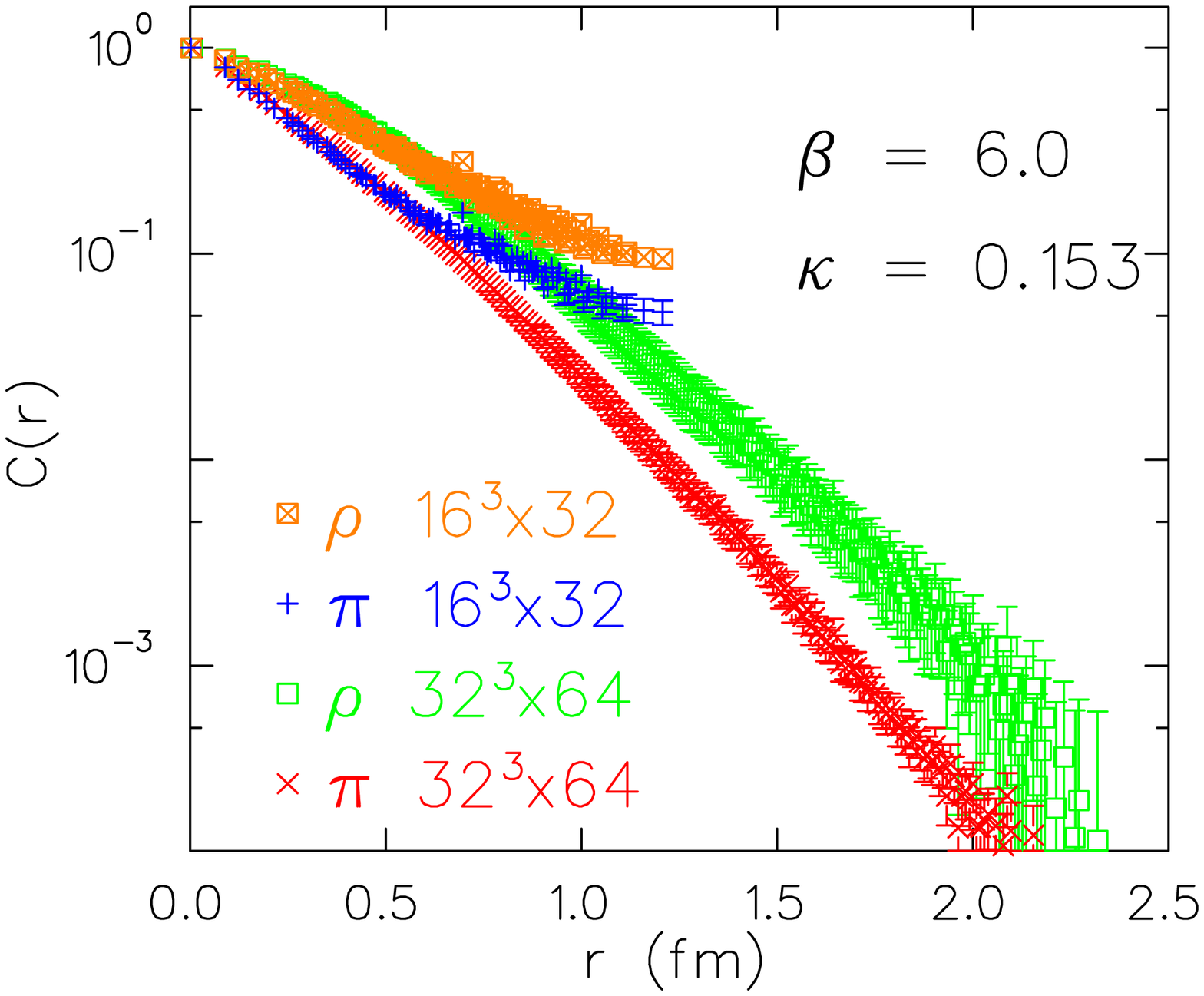,scale=0.3}
\caption{The charge density distribution for the pion and the rho mesons  
on lattices of size $16^3\times32$ and $32^3\times64$.}
\label{Fig:StandardCorr1}
\end{center}
\end{minipage}
\hfill
\begin{minipage}[t]{0.45\linewidth}
\begin{center}
\epsfig{file=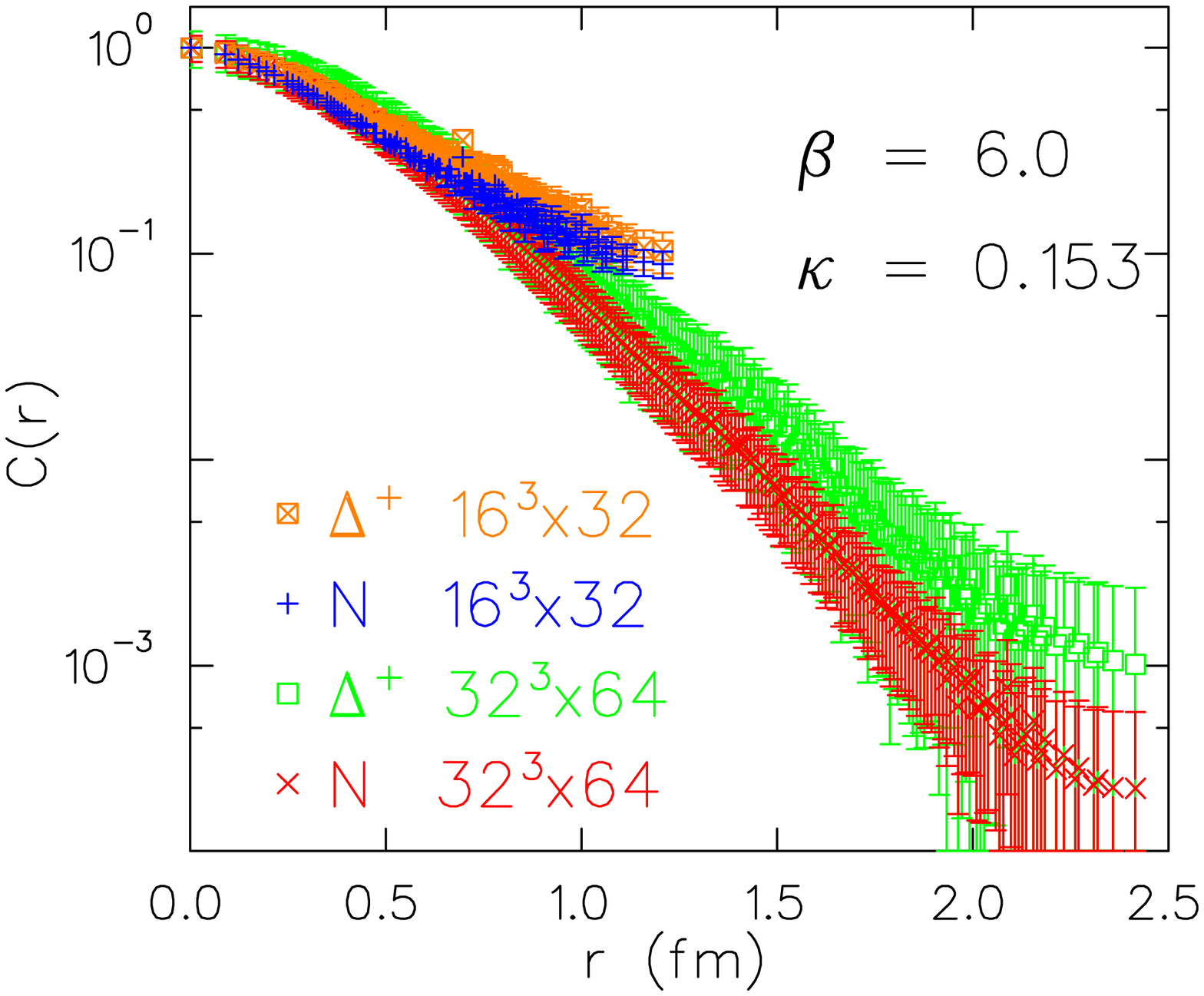, scale=0.3}
\caption{The charge density distribution for the nucleon and the $\Delta$ baryons
on lattices of size $16^3\times32$ and $32^3\times64$.}
\label{Fig:StandardCorr2}
\end{center}
\end{minipage}
\end{figure}

We follow the method used in ref. \cite{PRD:66-094503,*PRD:68-074504,*NPPS:128-1} to calculate density-density correlators using
point source propagators on a $32^3\times 64$ size lattice and compare with results obtained on a $16^3\times 32$ size lattice
\cite{PRD:66-094503,*PRD:68-074504,*NPPS:128-1} to check for finite volume effects. 
Since we do not sum over the sink the zero momentum projection is not carried out. 
For the lattice of size $32^3\times 64$ we use 100 quenched configurations while for the smaller lattice we use 220 quenched configurations both generated with the
Wilson action at $\beta=6.0$. We consider a value of $\kappa=0.153$ which gives
a pion to rho mass ratio, $m_\pi/m_\rho=0.84$. We compute the density correlation function for the pion,
the rho, the nucleon and the $\Delta$. The results shown in Figs.~\ref{Fig:StandardCorr1} and \ref{Fig:StandardCorr2} show that finite volume effects are large near the edges 
of small lattice. A  careful study of deformation on the large lattice is underway and will be reported elsewhere.
\vspace*{-0.75cm}
\section{Density-density correlators with zero momentum projection}
\begin{figure}[h]
\begin{minipage}[t]{0.3\linewidth}
\epsfig{file=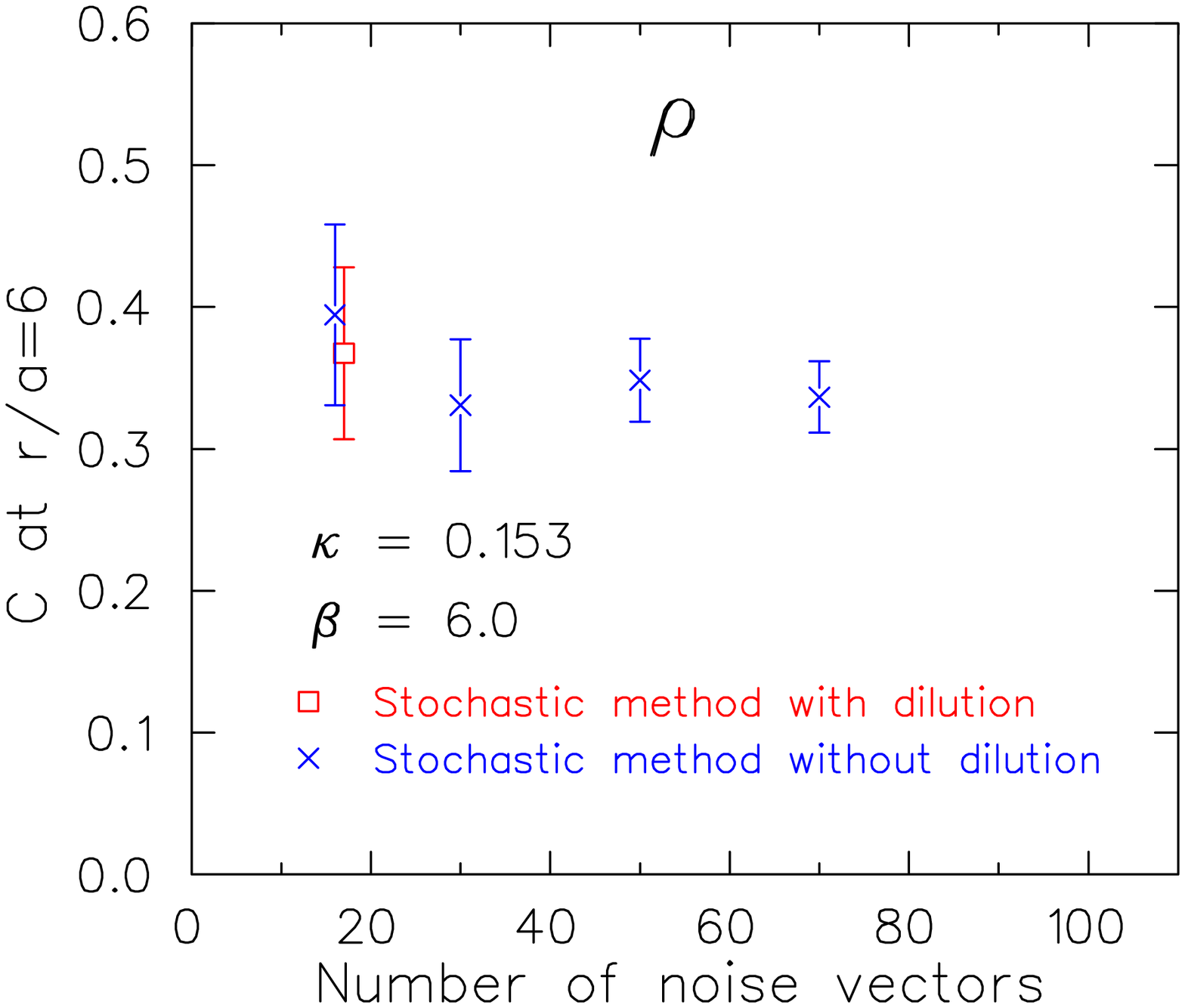,scale=0.25}
\caption{The rho density-density correlator versus the number of noise vectors. The red point (box) has been computed using dilution.}
\label{Fig:DiluteRho}
\end{minipage}
\hspace*{0.02\linewidth}
\begin{minipage}[t]{0.3\linewidth}
\centering
\epsfig{file=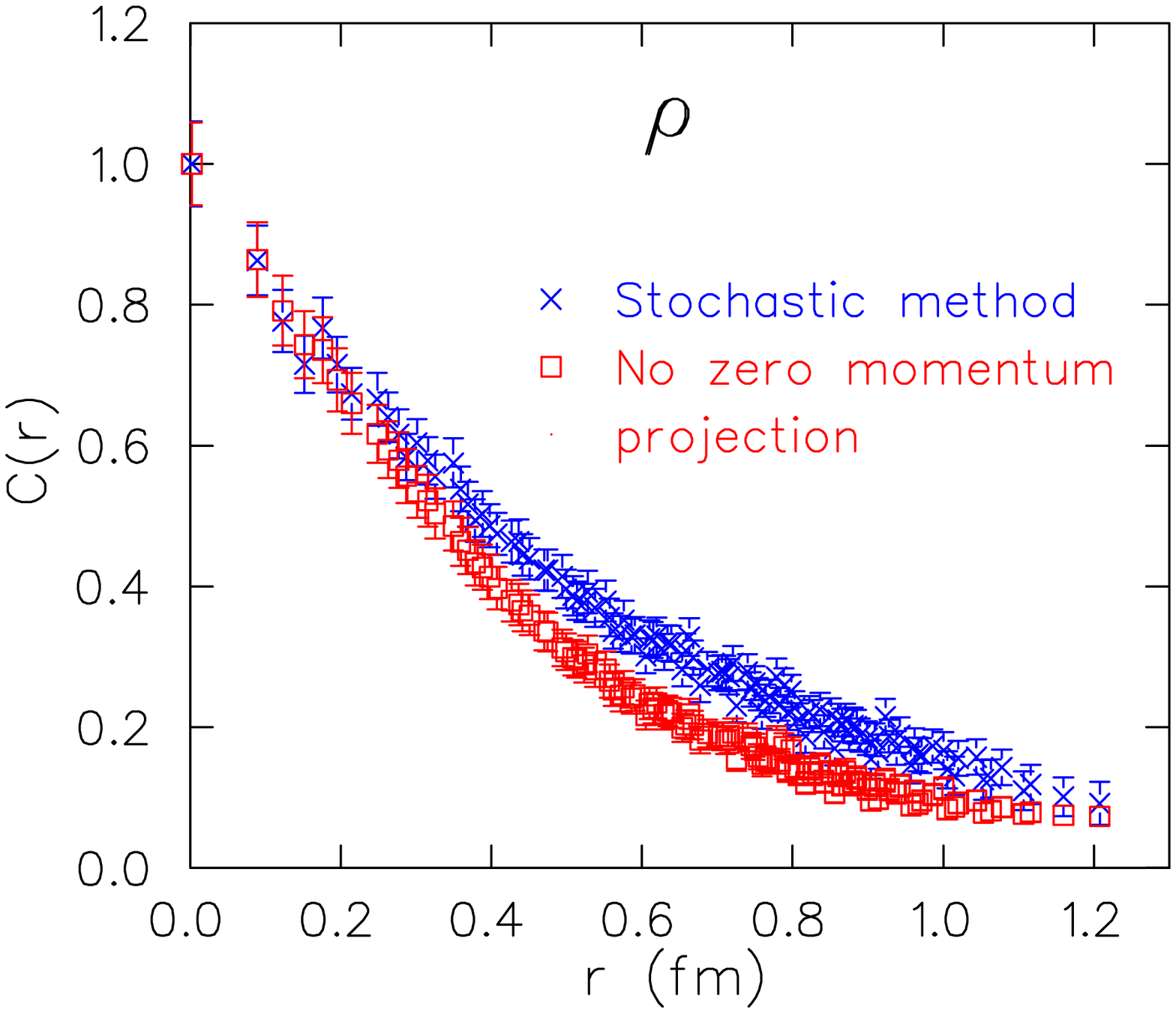,scale=0.25}
\caption{The charge density distribution of the rho
meson computed with no zero momentum projection (red squares) and with the stochastic method (blue crosses).}
\label{Fig:StandardStoch}
\end{minipage}
\hspace*{0.02\linewidth}
\begin{minipage}[t]{0.3\linewidth}
\centering
\epsfig{file=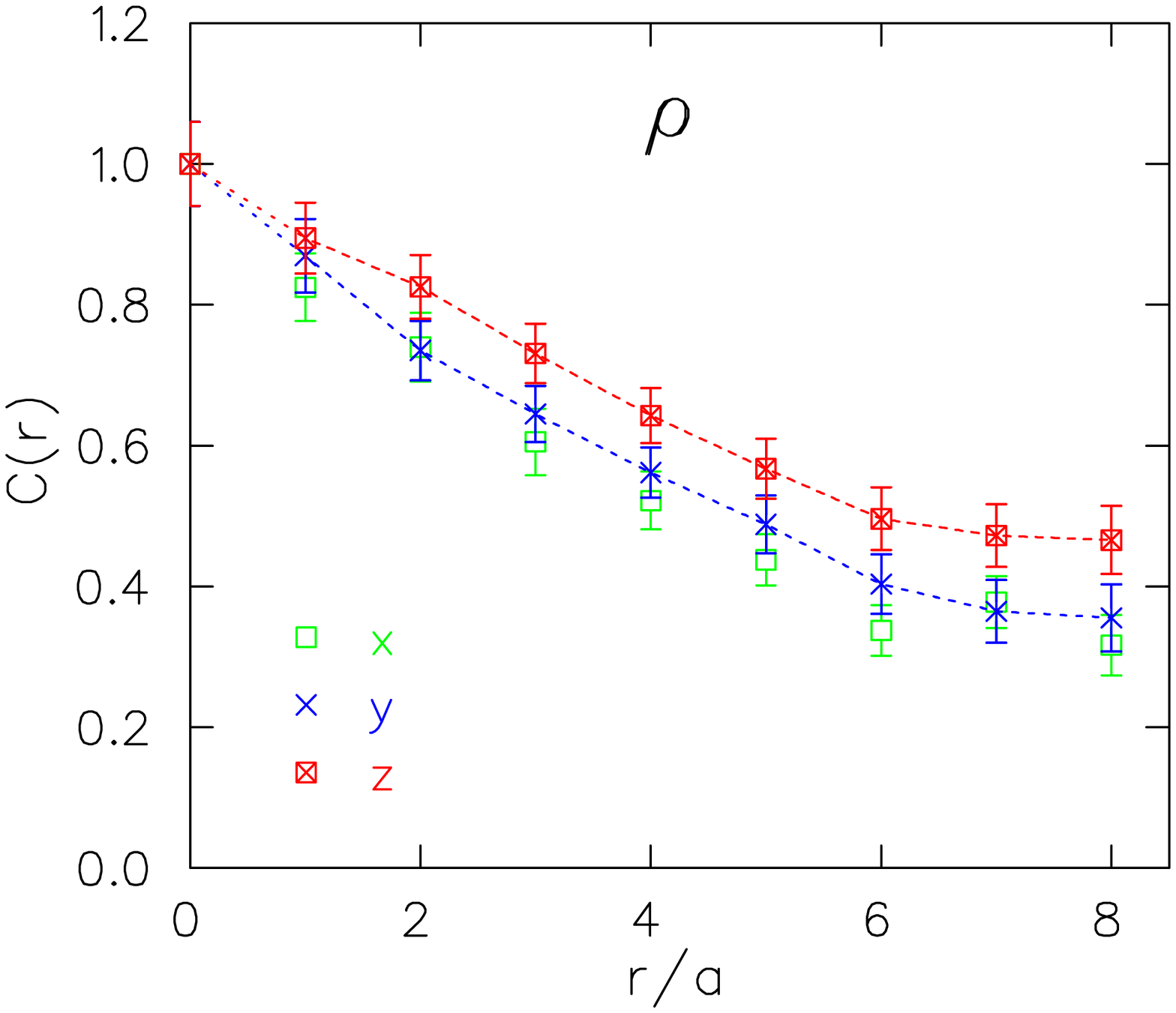,scale=0.25}
\caption{The charge density distribution of the rho
meson along the $x$, $y$ and $z$ axes separately.
The lines joining the points are to guide the
eye.}
\label{Fig:AsymmetryStoch}
\end{minipage}
\end{figure}

The stochastic propagator gives an estimate for the all-to-all propagator using
noisy sources \cite{PRD:58-034506,*PRD:59-074503,*HL:9408007}. The method assumes an ensemble of independently created noise
vectors with spin, color and space components randomly selected so that they obey the
conditions:
\be
\left<\eta_\mu^a\left(x\right)\eta^{b\dagger}_\nu\left(y\right)\right>=\delta\left(x-y\right)\delta_{\mu\nu}\delta_{ab}\qquad \left<\eta^a_\mu\left(x\right)\right>=0.
\label{Eq:Noise}
\ee
By solving the linear equation for each of the $r$ noise vectors in the ensemble we find:
\be
\eta^a_\mu\left(x\right)_r=M^{ab}_{\mu\nu}\left(y,x\right)\psi^b_\nu\left(x\right)_r\Rightarrow\left(M^{-1}\left(x,y\right)\right)^{ab}_{\nu\mu}
\eta^a_\mu\left(y\right)_r=\psi^b_\nu\left(x\right)_r
\ee
The all-to-all propagator can be constructed by taking the ensemble average of the
product between the solution vector and the noise source:
\be
\left<\psi^b_\nu(x)\eta^{a\dagger}_\mu(y)\right>=\left(M^{-1}(x,z)\right)^{bc}_{\nu\lambda}\left<\eta^c_\lambda(z)\eta^{a\dagger}_\mu(y)\right>=
\left(M^{-1}(x,y)\right)^{ba}_{\nu\mu}.
\ee
For the evaluation of density-density correlators we only need the spatial all-to-all propagator since the sink and density insertions are at fixed time slices. 
Whether we can in practice apply stochastic noise techniques to this problem 
depends upon how many noise vectors are needed to achieve convergence. We test the method by
considering the pion and the rho at $\beta=6.0$ and $\kappa=0.153$ on a lattice of size $16^3\times 32$.
We find that the error on $C(r)$ saturates for 30 noise vectors for the pion and 50 for the rho. 
In Fig.~\ref{Fig:DiluteRho} we show a representative case, namely $C(r)$ evaluated at $r/a=6$ for the rho as a function of the number of stochastic noise vectors using 
95 quenched configurations. In this work we adopt the saturation of the error as the criterion for convergence. 
In Fig.~\ref{Fig:StandardStoch} we compare the results obtained
with the stochastic method to those obtained without momentum projection for the same
number of configurations. We note that the statistical errors in both methods are the same
reflecting only gauge noise. The main difference seen in the projected zero momentum results is that the rho becomes broader. Such an
increase in the root mean square (r.m.s) radius of hadrons goes in the right direction since
typically the r.m.s radii were underestimated in previous evaluations within this
framework \cite{PRD:66-094503,*PRD:68-074504,*NPPS:128-1}. In Fig.~\ref{Fig:AsymmetryStoch} we show that the $z-x$ and $z-y$ asymmetry exhibited by the rho with
$J_z=0$ remains.
\vspace*{-0.1cm}
\subsection{Stochastic method with dilution}
We use a variant of the stochastic method which dilutes the noise
vectors along the $z$-axis, i.e.\ each noise vector only has non-zero values on a
chosen $z$-slice. This means that for our lattices at least 16 inversions are
required for each configuration. The method has been used in ref. \cite{HL:0505023} where the dilution is
carried out on the time-axis. As we already noted, in our case the stochastic propagator is only
needed on two fixed time slices on which we took independent noise vectors. 
Comparing in Fig.~\ref{Fig:DiluteRho} the density correlator for the rho with and without
dilution using the same number of noise vectors i.e.\ 16 noise vectors we find consistent results. 
Implementing dilution in the $z$-direction does not produce significant improvement and therefore we will not use it in what follows.
\vspace*{-0.5cm}
\subsection{Sequential stochastic method}
Instead of using two sets of noise vectors, one 
at $T/4$, the other at $T/2$, one can employ a sequential method, where the
all-to-all propagator with the noisy source at $T/4$ is used as the source for a
second inversion. Since we discard the noise at $T/2$ we expect that the stochastic
noise will be reduced.
Again we test this method for Wilson fermions at $\beta=6.0$ in the quenched theory taking $\kappa=0.153$ on a
$16^3\times 32$ lattice.

\begin{figure}[h]
\begin{minipage}[t]{0.3\linewidth}
\centering
\epsfig{file=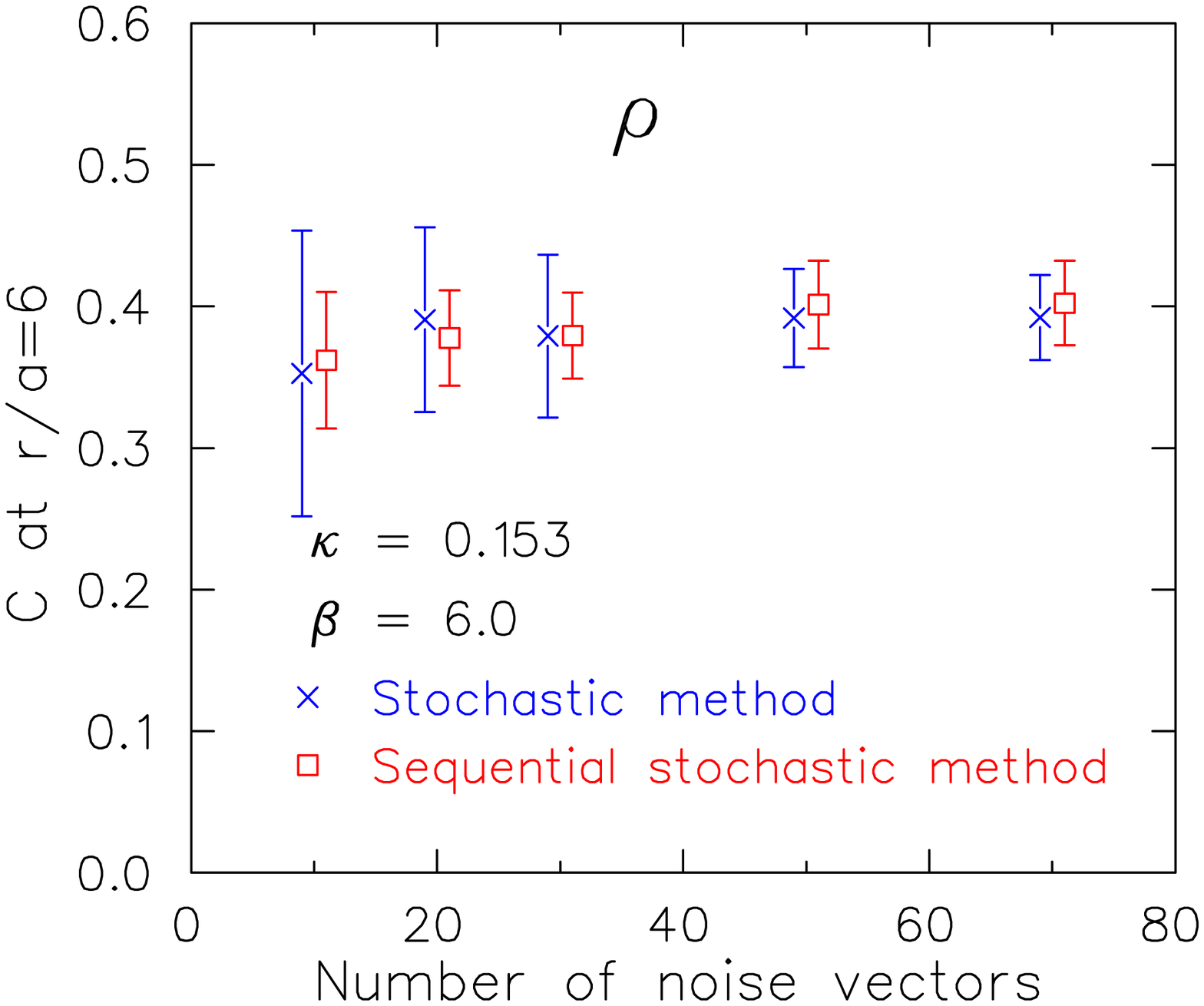,scale=0.25}
\caption{The density-density correlator at $r/a=6$ versus the number of
noise vectors needed for convergence
when two stochastic propagators are computed (blue
crosses) and when the sequential stochastic
method (red squares) is used.}
\label{Fig:NStochSeq}
\end{minipage}
\hfill
\begin{minipage}[t]{0.3\linewidth}
\centering
\epsfig{file=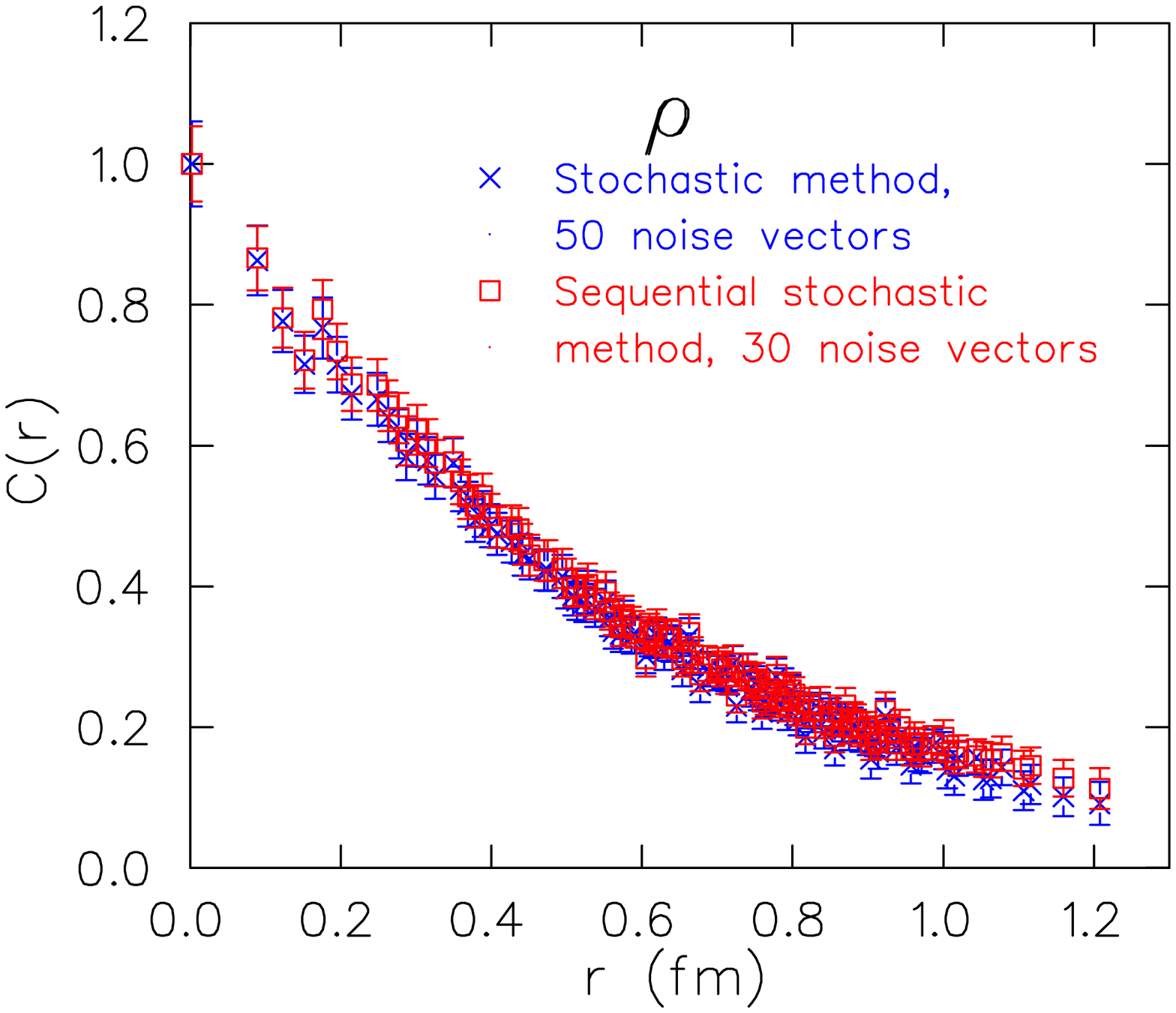,scale=0.25}
\caption{The charge density distribution of the rho
meson using the stochastic method (blue
crosses) and the sequential stochastic
method (red squares).}
\label{Fig:CorrStochSeq}
\end{minipage}
\hfill
\begin{minipage}[t]{0.3\linewidth}
\centering
\epsfig{file=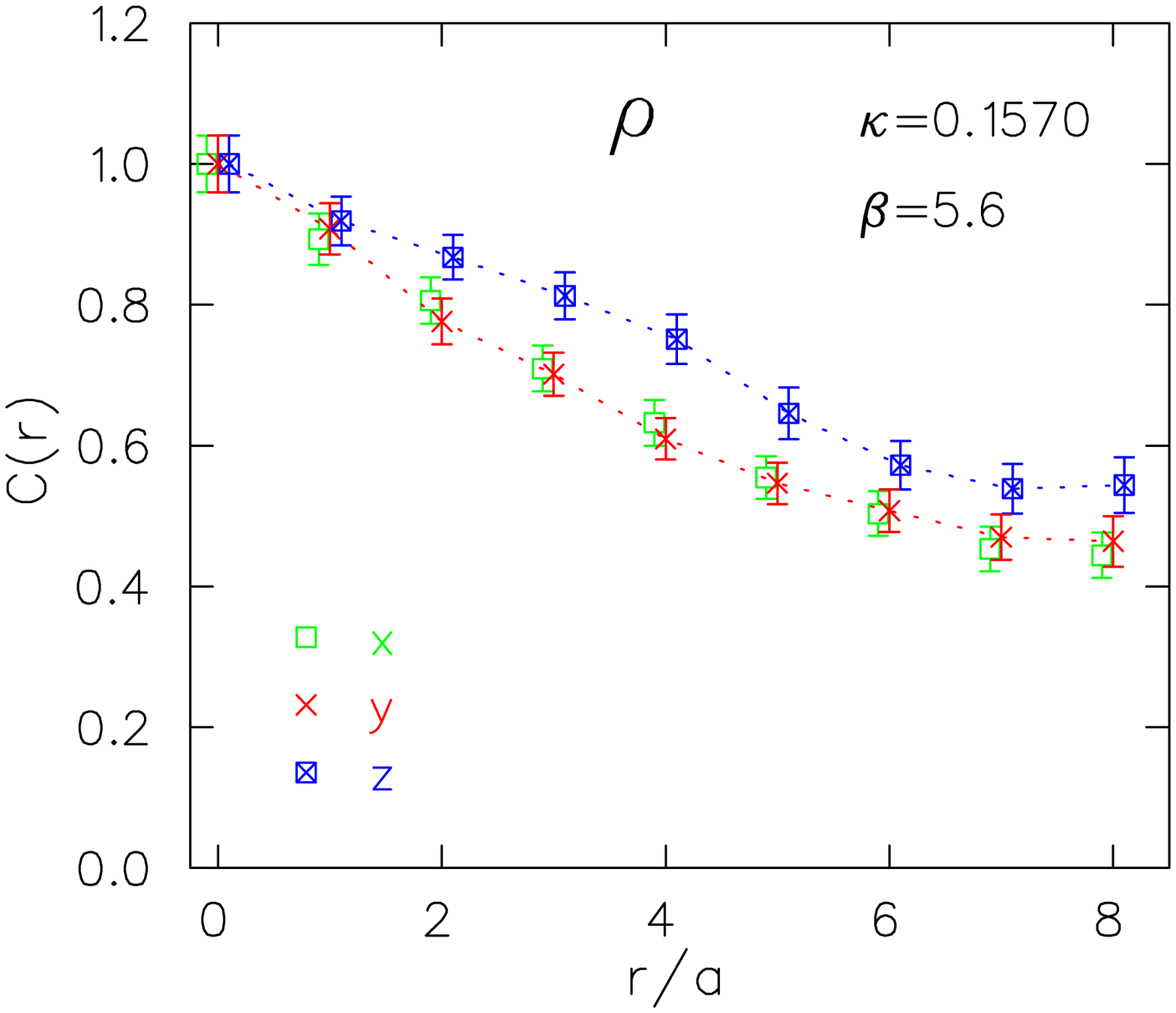,scale=0.25}
\caption{$C(x,0,0)$, $C(0,y,0)$ and $C(0,0,z)$ in full QCD at $\kappa=0.1570$
using 195 unquenched gauge fields (Symbols are
shifted for clarity).}
\label{Fig:AsymmUnqu}
\end{minipage}
\end{figure}

The density-density correlator we compute takes the form:
\be
C\left(\vec{y}\right)=-\sum_{\vec{x},\vec{z}}Tr\left[S_u\left(\vec{z}-\vec{x}\right)\gamma_0S_u\left(\vec{x}\right)\Gamma\gamma_5S_d^\dagger\left(\vec{x}+\vec{y}\right)\gamma_5\gamma_0S_d\left(\vec{x}+\vec{y}-\vec{z}\right)\Gamma\right]
\ee
where $\Gamma=\gamma_5$ for the pion and $\Gamma=\gamma_k$ for the rho $\left(k=1,2,3\right)$. By writing the all-to-all propagator as the product between the solution 
and noise vectors, we obtain
\be
C\left(\vec{y}\right)=-\frac{1}{N_r}\sum_{\vec{x},\vec{z},r}Tr\left[\gamma_0S_u\left(\vec{x}\right)\Gamma\gamma_5S_d^\dagger\left(\vec{x}+\vec{y}\right)\gamma_5\gamma_0S_d\left(\vec{x}+\vec{y}-\vec{z}\right)\Gamma\psi_r\left(\vec{z}\right)\eta_r^\dagger\left(\vec{x}\right)\right]
\ee
where the summation over the $z$ coordinate can be simultaneously carried out by solving the equation:
\be
\upsilon_r\left(\vec{x}+\vec{y}\right)=\sum_{\vec{z}} M^{-1}\left(\vec{x}+\vec{y},\vec{z}\right)\Gamma\psi_r\left(\vec{z}\right)
\ee
As can be seen in Fig.~\ref{Fig:NStochSeq} the sequential stochastic method converges when using 30
noise vectors per gauge field for the rho meson as compared to 50 needed when two stochastic propagators are computed. 
This results in a 40\% CPU time gain. The rho correlation
function using the two methods are compared in Fig.~\ref{Fig:CorrStochSeq} confirming that the results obtained with about half the noise estimators in the sequential 
stochastic approach  are consistent with those obtained using two stochastic propagators.
\vspace*{-0.5cm}
\section{Unquenched results}
Having tested the sequential stochastic method in the quenched theory we evaluate the
meson correlators using unquenched SESAM gauge configurations on a lattice of size
$16^3\times 32$. The rho asymmetry is shown in Fig.~\ref{Fig:AsymmUnqu} for 195 configurations at $\beta=5.6$ and
$\kappa=0.1570$, which gives $m_\pi/m_\rho=0.76$. We use spin, color and even-odd diluted noise
vectors \cite{HL:0505023}. We find convergence with 2 sets of noise vectors, i.e.\ with a total of $\left(4\times 3\times 2\right)\times 2 =48$ inversions per gauge field.
Wuppertal smearing is applied at the sink. Furthermore the point source propagators are calculated
on 9 sites to improve statistics. The unquenched results show a clear asymmetry in the rho
which is larger than what is observed in the quenched case at somewhat heavier pion mass.
\section{Conclusions}
We have explored the application of stochastic methods in the evaluation of density-density correlators. 
The all-to-all propagators are needed to project out zero momentum hadronic states.
Using the all-to-all propagators obtained with the standard stochastic technique at two
fixed time slices the pion and rho correlators are computed to a good accuracy. They
yield errors comparable to the evaluation of these correlators using point-to-all
propagators without zero momentum projection. Employing sequential all-to-all propagators reduces the cost by 40\%.
This also reduces memory requirements which is an important issue for the calculation of baryonic correlators.
Using unquenched SESAM configurations we clearly see that the rho is elongated along
its spin axis. The deformation observed increases as compared to the quenched case which
confirms our previous findings. Therefore it is an interesting question to see whether the $\Delta$
will develop a deformation in the unquenched theory at smaller quark masses. Work is in
progress for the calculation of the baryon density distributions using the stochastic
techniques described here.
\vspace*{-0.14cm}

\providecommand{\href}[2]{#2}\begingroup\raggedright\endgroup

\end{document}